\documentclass[prl,floatfix,twocolumn,showpacs,amsmath,amssymb]{revtex4}
\usepackage{graphicx,color}% Include figure files
\usepackage{dcolumn}% Align table columns on decimal point
\usepackage{bm}% bold math
\begin{document}

\title{A chiral spin liquid wave function and the Lieb-Schulz-Mattis theorem} 
\author{Sandro Sorella,$^{1}$ Luca Capriotti,$^{2}$ Federico Becca$,^{1}$ and
Alberto Parola$^{3}$}
\affiliation{
${^1}$ INFM-Democritos, National Simulation Centre, and SISSA, I-34014 Trieste, Italy \\
${^2}$  Kavli Institute for Theoretical Physics, University of California, Santa Barbara CA 93106-4030 \\
${^3}$ Istituto Nazionale per la Fisica della Materia and Dipartimento di Scienze, Universit\`a dell'Insubria, I-22100 Como, Italy 
}

\date{\today}

\begin{abstract}
We study  a chiral spin liquid wave function defined 
as a Gutwziller projected BCS state with a complex pairing 
function. 
After projection, spontaneous dimerization is found for any odd but  
finite number of chains, thus satisfying the Lieb-Schultz-Mattis theorem,
whereas for even number of chains there is no dimerization.
The two-dimensional thermodynamic limit is consistently reached for large
number of chains since the dimer order parameter vanishes
in this limit. This property clearly supports the possibility 
of a spin liquid ground state in two dimensions 
with a gap to all {\em physical} excitations and with no broken translation 
symmetry. 
\end{abstract}

%%%%%%%%%%%%%%%%%%
%%%%%%%%% PACS USED
%%%%  71.10.Fd Lattice fermion models (Hubbard model, etc.)
%%%%  74.20.Mn Nonconventional mechanisms (spin fluctuations, polarons and bipolarons, resonating valence bond model, anyon mechanism, marginal Fermi liquid, Luttinger liquid,
%%%%  71.10.Pm Fermions in reduced dimensions
%%%%  71.27.+a Strongly correlated electron systems; heavy fermions
%%%%%%%%%%%%%%%%%%

\maketitle

A long time after its first proposal,~\cite{anderson}
the existence of a spin liquid ground state (GS) in two-dimensional 
(2D) quantum spin one-half models is still a very controversial issue.
This is mainly because all one-dimensional (1D) or quasi-1D spin 
models that can be solved exactly, either numerically
or analytically,~\cite{mygosh,j1j2chain,twoleg} 
display a gap to the spin excitations only when a broken translation
symmetry is found in GS (e.g., in spin-Peierls 
systems~\cite{mygosh,j1j2chain})
or when the unit cell contains an even number of spin 1/2 
electrons (e.g., in the two-chain Heisenberg model~\cite{twoleg}).
Hence, in these models the electronic correlations do not play a crucial role
since their GS can be adiabatically connected
to  a band insulator without any transition. 
This important property of insulators, which clearly holds in 1D systems, 
has been speculated to be generally valid even in higher dimensions, 
as it appears to follow from a general result,
the Lieb-Schulz-Mattis (LSM) theorem,~\cite{lsm} whose range of validity 
has been extended to more interesting 2D cases.~\cite{lsma,oshikawa}  

Recently, there has been an intense theoretical and
numerical investigation of non-magnetic wave functions obtained after 
Gutzwiller projection of the GS of a BCS 
Hamiltonian.~\cite{rvb,rainbow,ivanov,arun} 
On a rectangular $L_x\times L_y$ lattice, this state 
can be written in the following general form:
\begin{equation} \label{wf}
|p{-}{\rm BCS} \rangle  = \hat{P}_G  \left[\sum_k f_k \hat{c}^\dag_{k,\uparrow}
\hat{c}^\dag_{-k,\downarrow} \right]^{\frac{N}{2}} |0\rangle~, 
\end{equation}
where $N$ is the number of electrons 
(equal to the number of sites, {\em i.e.}, $N=L_x\times L_y$), $\hat{P}_G$ 
is the Gutzwiller projection onto the subspace of no doubly 
occupied sites, and $\hat{c}^\dag_{k,\uparrow}$ and 
$\hat{c}^\dag_{k,\downarrow}$ 
are creation operators of a spin up or a spin down 
electron, respectively. These are defined in a plane-wave state 
with momentum $k$ allowed by the chosen boundary conditions: 
periodic (PBC) or antiperiodic (APBC) in each direction.  
It is worth noting that, after the projection, the wave function (\ref{wf})
corresponds to PBC on the spin Hamiltonian, regardless the choice of
boundary conditions on the electronic states.
The pairing function $f_k$ can be easily related to the gap 
function $\Delta_k$ and the bare dispersion 
$\epsilon_k=-2(\cos k_x + \cos k_y)$ of the BCS Hamiltonian:
\begin{equation} \label{hbcs}
\hat{H}_{\rm BCS}=\sum_{k,\sigma} 
\epsilon_k \hat{c}^\dag_{k,\sigma} \hat{c}_{k,\sigma} + 
\sum_k \left( \Delta_k \hat{c}^\dag_{k,\uparrow} \hat{c}^\dag_{-k,\downarrow}
+{\rm h.c.} \right)~, 
\end{equation}
by means of the simple relation 
$f_k = \Delta_k /(\epsilon_k + E_k)$, 
where the gap function $\Delta_k$ can be in general any complex function
even under inversion ($\Delta_k=\Delta_{-k}$), as 
required here for a singlet wave function,~\cite{rainbow}
and $E_k=\sqrt{|\Delta_k|^2+\epsilon_k^2}$ 
represents the spin-half excitation energies of $H_{\rm BCS}$.

As clearly pointed out by Wen,~\cite{wen}
in presence of a finite gap $\Delta_{BCS}$ in the thermodynamic limit,
such that $E_k\ge \Delta_{BCS}>0$, the corresponding BCS finite correlation 
length  is expected to be robust under Gutzwiller 
projection.
Here, we restrict to this class of non-magnetic states, considering
the projected BCS ($p$-BCS) state which is obtained 
by a $d+id$ gap function of the following form:
\begin{equation}\label{gap}
\Delta_k =\Delta_{x^2-y^2}  (\cos k_x - \cos k_y) +
i \Delta_{xy} \sin k_x \sin k_y~.
\end{equation}
This wave function breaks the time reversal symmetry $\hat{T}$ and the parity
symmetry ($x \leftrightarrow   y$  in 2D) $\hat{P}$, whereas 
$\hat{T} \bigotimes \hat{P}$ is instead a well defined symmetry. 
Hence, this spin liquid wave function may have a non-vanishing value of the 
so-called {\em chiral} order parameter:~\cite{wilcek} 
\begin{equation} \label{chiral}
\hat{O}_{\rm C} = \frac{1}{N} \sum_{i} \hat{ \bf S}_i  \cdot 
(\hat{ \bf S}_{i+d_x} \times \hat{ \bf S}_{i+d_y})~, 
\end{equation}
with $d_x=(1,0)$ and $d_y=(0,1)$.
Chiral spin liquids were introduced a long time ago,~\cite{wilcek,kalmaier}
however, to our knowledge, this is the first attempt to represent this
class of states in the $p$-BCS framework.
On finite size systems, we take the real part of 
the complex wave function (\ref{wf}), so that all the finite-size symmetries, 
including parity, are satisfied and a spontaneously broken lattice symmetry  
can occur only in the thermodynamic limit.  
 
In the following we consider in more detail the relation of the $p$-BCS 
wave functions with the LSM theorem.
Given a short-range spin Hamiltonian $\hat{H}$, on a $L_x \times L_y$
rectangle (PBC on the $x$ direction are assumed) and 
a variational state $|\psi_0\rangle$ with given momentum,
we can define another variational state, 
$|\psi_0^\prime\rangle =\hat{O}_{LSM}|\psi_0\rangle$,
by means of the LSM operator:
\begin{equation} \label{lsm}
\hat{O}_{LSM} = \exp \left[ i \sum_r \frac{ 2 \pi x}{L_x}\hat{S}^z_r  \right]~,
\end{equation}
where $r=(x,y)$ indicates the position of each site on the lattice. 
The new variational state $|\psi^\prime_0\rangle$ has
the following properties:~\cite{lsma}
\begin{enumerate}
\item Its energy expectation value 
differs at most by $O(L_y/L_x)$ from the variational 
energy of $|\psi_0\rangle$.
\item If $L_y$  is odd, regardless of the boundary conditions on the 
$y$ direction, the momenta parallel to $x$ 
corresponding to $|\psi_0\rangle$ and $|\psi^\prime_0\rangle$ 
differ by $\pi$. Hence: $\langle \psi^\prime_0 | \psi_0 \rangle =0$. 
\end{enumerate}

For 1D or quasi-1D system with odd number of chains $L_y$ and 
vanishing aspect ratio ($L_y/L_x \to 0$ for $L_x\to \infty$), 
by applying the LSM operator to the actual GS, 
it is possible to construct an excitation of the system 
with momentum $(\pi,0)$ which becomes degenerate with the GS
in the thermodynamic limit. This implies in turn either a gapless spectrum
or, in presence of a finite gap, a two-fold degenerate  GS
with a doubling of the unit cell and a spontaneously broken 
translation symmetry.
For instance, the presence of a singlet zero-energy excitation with
momentum $(\pi,0)$ is just a characteristic of spontaneous 
spin-Peierls dimerization, as it appears for example in the 
Majumdar-Gosh chain.~\cite{mygosh} This result, holding rigorously
in the limit of vanishing aspect ratio, has been argued to apply
in general for 2D systems.~\cite{lsma}
In the following, we will show, with an explicit example,
that this result in 2D does not necessarily  
imply spontaneous dimerization, but 
topological degeneracy of the GS.

It is simple to show that
$\hat{O}_{LSM} |p{-}{\rm BCS}\rangle = |p{-}{\rm BCS}^{\prime}\rangle$,
namely {\em the same type of wave function} of Eq.~(\ref{wf})
is obtained, with the changes below:
\begin{eqnarray} \label{changervb}
k & \to & \bar k~,   \label{kbar}  \\
f_{k} & \to & \bar f_{\bar k}=f_{\bar k - (\pi/L_x,0)}=f_{k}~,  \label{fk}
\end{eqnarray}
where the new quantized momenta $\bar k=k+(\pi/L_x,0)$
are obtained by interchanging PBC with APBC
in the $x$ direction only and Eq.~(\ref{fk}) means that the
pairing function is calculated with the old momenta $k$:
$f_k \hat{c}^\dag_{\bar k,\uparrow} \hat{c}^\dag_{-\bar k,\downarrow}$.
By definition, the wave function $|p{-}{\rm BCS}^\prime\rangle$ 
has therefore the same quantum numbers predicted by the LSM theorem, 
the change of momentum being implied by Eq.~(\ref{kbar}).
The reason why the momentum of the wavefunction (\ref{wf}) can be non zero 
for odd number of chains is indeed rather subtle but easy to verify.
Indeed the $x-$translation operator with APBC 
translates all creation operators, but the ones 
belonging to the boundary are also multiplied by a phase factor $(-1)$.  
This translation operator always leaves invariant the 
$|p{-}{\rm BCS}^\prime\rangle$ wave function.
However, for a spin state with one electron per site each 
configuration has always $L_y$ electrons at the boundary, so that the physical 
spin translation operation (defined with PBC), differs  from 
the APBC one for an overall 
phase $(-1)^{L_y}$, namely a momentum $(\pi,0)$ for odd number of chains.
Analogously, the excitations obtained by modifying only the boundary 
conditions in the BCS Hamiltonian (in the $x$ and/or $y$ direction), 
namely using Eq.~(\ref{kbar}) (and/or its equivalent for the $y$ direction)
and $f_{k} \to f_{\bar k}$,  
may display in 2D the topological degeneracy of this 
spin liquid wave function.~\cite{ivanov,arun} 

\begin{figure}
\includegraphics[width=0.45\textwidth]{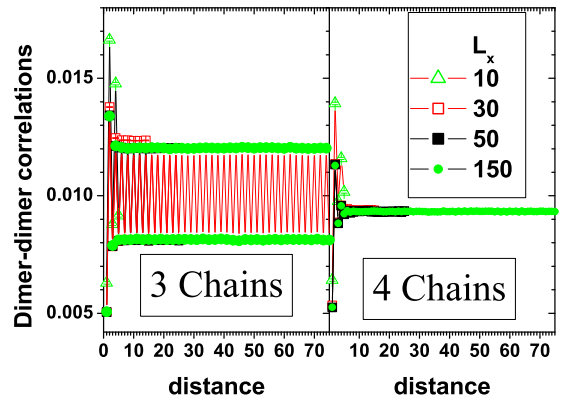}
\vspace{-5mm}
\caption{\label{dimerchiral}
Dimer-dimer correlation functions as a function of the distance for three 
and four chains with PBC on both directions, using the wave function 
(\protect\ref{wf}) with $\Delta_{x^2-y^2}=2$ and $\Delta_{xy}=1$.
}
\end{figure}

\begin{figure}
\includegraphics[width=0.45\textwidth]{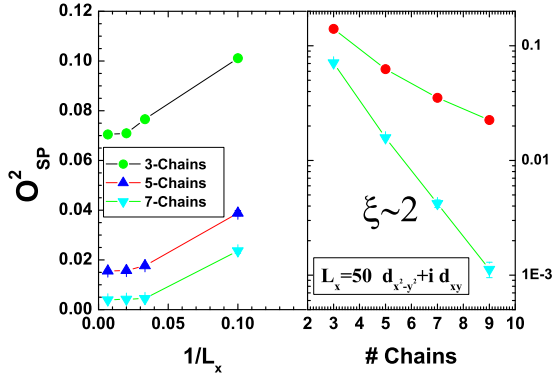}
\vspace{-5mm} 
\caption{\label{dimchain}
Left: Finite-size scaling of the dimer order parameter in the 
rectangular geometries with an odd number of chains. 
Right: Semi-log plot of the dimer order parameter 
as a function of the number of chains.
In both plots, the data refer to the chiral wave function  
with $\Delta_{x^2-y^2}=2$ and $\Delta_{xy}=1$, whereas full dots in the 
right panel refer to the short-range RVB (the latter wave function 
having $\xi \to \infty$ in 2D).
}
\end{figure}

\begin{figure}
\includegraphics[width=0.45\textwidth]{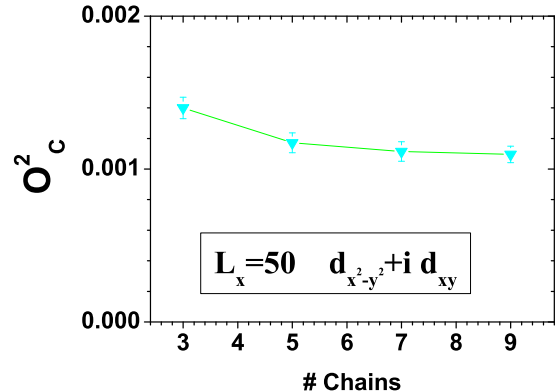}
\vspace{-5mm}
\caption{\label{chiralfig}
Chiral order parameter as a function of the number of chains.
Its value is determined by studying the chiral-chiral correlation
functions on the smallest parallel triangles at the maximum distance along
the long direction. $L_x=50$ is indeed large enough for converged results at
fixed number of chains, as the wave function has a large gap $\Delta_{BCS}=1$.
The variational parameters are $\Delta_{x^2-y^2}=2$ and $\Delta_{xy}=1$.
}
\end{figure}

We now  assume that the the $p$-BCS wave function (\ref{wf}) with
the $\Delta_k$ given by Eq.~(\ref{gap}) represents   
the GS of some short-range Hamiltonian. 
However, here we do not address the question whether this  
chiral wave function can be stabilized in some physical Hamiltonian. 
Certainly explicit Hamiltonians with short-range off-diagonal matrix elements 
can be constructed by a simple inversion problem scheme.~\cite{effective} 
Having a finite gap $\Delta_{BCS}$, 
this wave function describes a spin system with a finite correlation length,
and consequently a finite triplet gap. 
The LSM theorem can be then applied in the geometries where 
it holds, like, for instance, the three-leg ladder with PBC in 
both directions.
As shown in Fig.~(\ref{dimerchiral}), it is clear 
that spontaneous dimerization is obtained in the 
thermodynamic system for this geometry, as
the dimer-dimer correlation functions on each chain,
\begin{equation}\label{dimerdimer}
\Delta(r-r^\prime)
=\langle \hat{S}_r^z\hat{S}_{r+d_x}^z \hat{S}_{r^\prime}^z 
\hat{S}_{r^\prime+d_x}^z\rangle~, 
\end{equation}
behaves  for large distance
as  $(-1)^{(x-x^\prime)} O_{\rm SP}^2/36+{\rm const.}$,
being $O_{\rm SP}$ the dimer order parameter.~\cite{j1j2chain,thepig}
In strong analogy with the one-dimensional Heisenberg 
chain in the gapped phase,
the broken translation symmetry allows the system to satisfy the 
LSM theorem. In fact, it implies two degenerate singlet 
states with momentum differing by $(\pi,0)$. 
In contrast, on any even-leg ladder, where the LSM
theorem does not imply the degeneracy, the $p$-BCS state does not break 
translational invariance, as illustrated in Fig.~\ref{dimerchiral} for the
four-leg system.
Despite the dichotomy between the odd and even chain cases,
the 2D thermodynamic limit can be still consistently defined.
In fact, as it is clearly shown in Fig.~\ref{dimchain}, though a finite 
dimer order parameter is obtained for any odd 
chain ladder, the order parameter is {\it exponentially}
decreasing with the number of chains [see Fig.~\ref{dimchain}].  
This implies that the broken symmetry, 
which is correctly obtained for odd but finite number of 
chains, represents an irrelevant effect in the 2D thermodynamic system.
Nonetheless, in the 2D system, the GS can possess degenerate topological 
excitations. For instance, the matrix element of the
dimer operator with momentum $(\pi,0)$ between the two degenarate singlet 
states, which is finite on any finite number of chains, 
decreases exponentially with increasing $L_y$, as it is bounded by order
parameter (see Fig.~\ref{dimchain}).
Remarkably, the chiral order parameter remains instead 
a genuine feature of this
variational wave function even in the 2D limit, as shown in 
Fig.~\ref{chiralfig}, where an order parameter 
$O_{\rm C}=\sqrt{\langle \hat{O}^2_{\rm C} \rangle} \simeq 0.03$ was found for 
this wave function. 

We have given here  a clear example that a spin liquid GS
can be stable in 2D, and  yet satisfying  all the known constraints 
given  by the LSM theorem.
Indeed, spontaneous broken translation symmetry is obtained for any 
odd number of chains, a remarkable feature since before projection 
the wave function is translational invariant. 
As the number of chains increases the chiral spin liquid appears 
only in 2D systems, where the spin-Peierls dimer order 
parameter converges to zero, and no broken translation symmetry is 
implied in the thermodynamic limit. 

\begin{figure}
\includegraphics[width=0.45\textwidth]{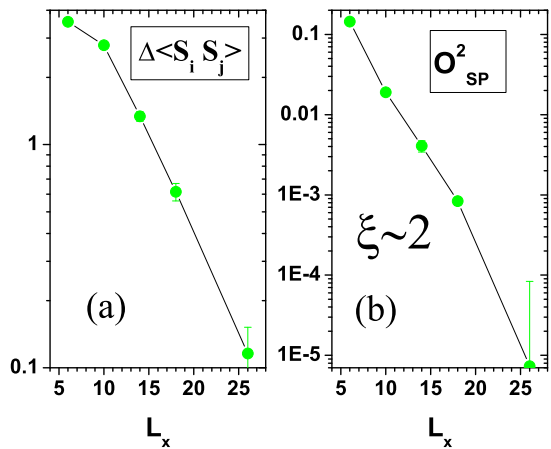}
\vspace{-5mm}
\caption{\label{deltae}
(a): Difference in the expectation value of the nearest-neighbor spin-spin
correlations multiplied by the number of sites
between the two orthogonal states with the same
quantum numbers predicted by the LSM theorem.
(b): Finite-size scaling of the dimer order parameter in rectangular lattices
with finite aspect ratio.
The variational parameters are $\Delta_{x^2-y^2}=2$ and $\Delta_{xy}=1$.
and $L_y = L_x+1$.
}
\end{figure}

The chiral spin-liquid described by the $p$-BCS wave function 
is also consistent with a recent extension of the LSM to 
2D systems with finite aspect ratio $L_y/L_x$. 
In this case, as pointed out by Oshikawa,~\cite{oshikawa} the state 
$|\psi_0^\prime\rangle=\hat{O}_{LSM}|\psi_0\rangle$, 
is not necessarily degenerate with the starting wave function 
$|\psi_0\rangle$, as
in the usual LSM construction. However, through a well-defined 
adiabatic evolution -- in analogy 
with Laughlin's treatment of the quantum Hall effect -- 
one can obtain a different 
state $|\tilde \psi^\prime_0\rangle$, with the same spatial quantum numbers of
$|\psi_0^\prime\rangle$ and degenerate with 
$|\psi_0\rangle$. 
If $|\psi_0\rangle$ is described by a $p$-BCS
state (\ref{wf}), it is easy, by a small change of the pairing function
of the state $|\psi_0^\prime\rangle$,
i.e., $\bar f_{\bar k} \to f_{\bar k}$, to define a state 
$|\tilde \psi^\prime_0\rangle$ with the same momentum implied by LSM theorem
but expected to be degenerate with $|\psi_0\rangle$.
In fact, the wave function $|\tilde \psi_0^\prime\rangle$ can be 
obtained with the {\em same} BCS Hamiltonian (\ref{hbcs}) 
with APBC in the $x$ direction. 
Then, in presence of a finite gap
in the excitation spectrum $\Delta_{BCS}>0$, the wave functions 
$|\psi_0\rangle$ and $|\tilde\psi_0^\prime\rangle$ 
have the same value on any physical operator,
and in particular the Hamiltonian, so that the considered 
states are degenerate in the thermodynamic limit.   
This is shown, for instance in Fig.~\ref{deltae}(a) 
for the nearest-neighbor total energy contribution.

The state $|\tilde\psi_0^\prime\rangle$,
obtained by the adiabatic evolution of the LSM excitation,
$|\psi_0^\prime\rangle=\hat{O}_{LSM}|\psi_0\rangle$, is no longer connected
to $|\psi_0\rangle$ by any physical operator, and, therefore,
no spontaneous dimerization is implied in the thermodynamic limit.
Indeed, as shown in Fig.~\ref{deltae}(b), also in geometries with 
non-zero aspect ratio the $p$-BCS wave function has a finite 
dimer correlation length. 
We have therefore shown a clear counter example to the so-called 
Oshikawa conjecture~\cite{oshikawa} 
that no spin liquid is possible in two or higher dimensions.
For the Hamiltonians having the wave function (\ref{wf}) as the unique 
GS,~\cite{effective} the above argumentations are 
{\em clear cut and conclusive}.  
 
In general, projected wave functions of the type (\ref{wf}),
with a finite gap $\Delta_{BCS}>0$, do not necessarily break 
time reversal, like, for instance, by using a chemical potential 
outside the band,~\cite{arun} $d+is$ symmetry,~\cite{wen} 
or more simply $s$ symmetry. 
For all these wave functions we expect a finite dimer order parameter 
in lattices with infinite aspect ratio in agreement with 
the LSM theorem.   

The finite dimer correlation length represents a remarkable property of 
this projected chiral BCS wave function.
For instance, the conventional short-range RVB~\cite{bonsteel} 
displays instead power-law dimer correlations in 2D,~\cite{fisher} 
and therefore gapless features which may describe 
a singular point rather than a 2D spin-liquid phase. 
The fundamental difference between our chiral RVB and the short-range 
one is due to the violation of the Marshall sign rule in the former case. 
 This property is observed for instance  
in the ground state of the  $J_1-J_2$ model in the strongly  frustrated 
region $J_2\simeq 0.5$.\cite{rainbow}   Thus   
we expect  that {\em a true spin liquid phase cannot be stabilized 
in particular spin models where
the wave function signs are not allowed to change as a function of the
parameters  of the Hamiltonian}.~\cite{sandvik}
The complex wave function with $d$-wave symmetry proposed here appears to be
a reasonable way to open a gap close to a gapless 
antiferromagnetic phase in a fully translationally invariant spin  
model that allows the violation of the Marshall sign rule, i.e. {\em generic} 
frustrated models on a square lattice.  
The chiral order parameter can be indeed small and vanishing for 
$\Delta_{xy} \to 0 $ when the correlation length $\xi \to \infty$. 
A chiral spin liquid of this form with a reasonable correlation length 
$\xi\simeq 20$ provides therefore  
an almost negligible chiral order parameter, which is extremely difficult
to detect numerically and may fit well 
with the present experimental resolutions in High temperature 
superconductors.~\cite{campuzano}   

In conclusion we have presented a clear example that a spin liquid 
GS with a gap to all physical excitations, though being with single 
electron per unit cell, can be realized without 
violating the LSM theorem and its generalizations in 2D.~\cite{oshikawa}
Our results provide a clear support to the possibility of a true 
Mott insulator at zero temperature, an insulator that cannot be adiabatically
continued to any band insulator, showing that  
the effect of correlation maybe highly non trivial in 2D systems.

One of us (S.S.) thanks R. Laughlin for many useful discussions and for 
particularly exciting comments. Thanks to D.J. Scalapino 
for useful discussions, and for his kind 
hospitality at UCSB (S.S.). 
This work was substantially  supported  by 
INFM-PRA-MALODI, and partially by MIUR-COFIN 01. L.C.
was supported by NSF under grant 000000000.

\end{document}